\documentclass[aps,prl,a4paper,tightenlines,superscriptaddress,twocolumn,floatfix,final]{revtex4}
\usepackage{latexsym}
\usepackage{amssymb}
\usepackage[dvips]{graphics}
\usepackage[pdftex]{graphicx}
\usepackage[latin1]{inputenc}
\usepackage{amsmath}
\usepackage{amsfonts}
\usepackage{amssymb}
\usepackage{amsthm}

\newcommand{\beq}{\begin{eqnarray}}
\newcommand{\eeq}{\end{eqnarray}}
\newcommand{\be}{\begin{eqnarray}}
\newcommand{\ee}{\end{eqnarray}}

\begin{document}

\title{Varying Alpha: New Constraints from Seasonal Variations}
\author{John D. Barrow}
\affiliation{DAMTP, Centre for Mathematical Sciences, University of Cambridge,
Wilberforce Road, Cambridge CB3 0WA, UK}
\author{Douglas J. Shaw}
\affiliation{School of Mathematical Sciences, Queen Mary, University of London, London E1
4NS, UK}

\begin{abstract}
We analyse the constraints obtained from new atomic clock data on the
possible time variation of the fine structure `constant' and the
electron-proton mass ratio and show how they are strengthened when the
seasonal variation of Sun's gravitational field at the Earth's surface is
taken into account. We compare these bounds with those obtainable from tests
of the Weak Equivalence Principle and high-redshift observations of quasar
absorption spectra consistent with time variations in the fine structure
constant.

\end{abstract}

\maketitle

General relativity and the standard model of particle physics depend on at
least 27 independent parameters which determine the relative strengths of
the different forces, matrix angles and phases, and the masses of all known
fundamental particles. These parameters are commonly referred to as the
fundamental constants of Nature and we are not able to explain or predict
any of their precise numerical values. This characterisation may ultimately
turn out to be a misnomer, as several modern proposals for the extension of
fundamental physics beyond the standard model predict that many of these
parameters are neither strictly fundamental nor constant. The precise values
that the `constants' take at a given instance are often
determined in terms of the vacuum expectation values of one or more scalar
fields and space-time variation of the `constants', at some level, is
therefore a prediction common to most of these theories of new physics. We
also recognize that if the true fundamental theory exists in more than four
spacetime dimensions then the most fundamental constants are defined there
and the four-dimensional `shadows' that we observe and call `constants of
Nature' can vary without undermining the status of constants in the
higher-dimensional theory. Moreover, slow changes in the mean size of these extra dimensions will cause dimensionless constants
to change at the same rate \cite{bmarc}. Consequently, observational
investigations of the constancy of these traditional constants may play an
important role in elucidating the properties of any 'Theory of Everything'.

Experimental and observational efforts to constrain the level of any
possible time variation in some of these constants have a history that
pre-dates the modern theories which predict how they might vary (for
overviews see Refs. \cite{revs}). Until recently, all observational studies
found no evidence for variations in non-gravitational constants. In the last
decade, however, data from a number of astronomical observations have
provided suggestions that at least two of these constants, the fine
structure constant: $\alpha = e^{2}/4\pi \epsilon _{0}\hbar c$, and the electron-proton mass
ratio: $\mu =m_{e}/m_{p}$ might have varied slightly over cosmological time.
Using a data set of 128 Keck-HIRES quasar absorption systems at redshifts $%
0.5<z<3$, and a new many-multiplet (MM) analysis of the line separations
between many pairs of atomic species possessing relativistic corrections to
their fine structure, Webb \emph{et al.} \cite{webb} found the observed
absorption spectra to be consistent with a shift in the fine structure
constant, $\alpha$, between those redshifts
and the present day, of $\Delta \alpha /\alpha \equiv \alpha (z)-\alpha
(0)/\alpha (0)=-0.57\pm 0.10\times 10^{-5}$. A smaller study of 23 VLT-UVES
absorption systems between $0.4\leq z\leq 2.3$ by Chand \emph{et al.} \cite%
{chand} initially found $\Delta \alpha /\alpha =-0.6\pm 0.6\times 10^{-6}$
by using an approximate version of the full MM technique. However a recent
reanalysis of the same data by Murphy \emph{et al.} using the full unbiased
MM method increased the uncertainties and suggested the revised figure of $%
\Delta \alpha /\alpha =-0.64\pm 0.36\times 10^{-5}$ for the same data \cite%
{murphyrev}.   These investigations rely on the statistical gain from large
samples of quasar absorption spectra. By contrast, probes of the
electron-proton mass ration can use single objects effectively. Reinhold 
\emph{et al.} \cite{reinhold} found a $3.5\sigma $ indication of a variation
in the electron-proton mass ratio $\mu $ $\equiv m_{e}/m_{pr}$over the last $%
12\,\mathrm{Gyrs}$: $\Delta \mu /\mu =(-24.4\pm 5.9)\times 10^{-6}$ from $%
H_{2}$ absorption in a different object at $z=2.8$. However, recently Murphy
et al \cite{murph mass} have exploited the $\mu $ sensitivity of ammonia
inversion transitions \cite{flam} compared to rotational transitions of CO,
HCN, and HCO$^{+}$ in the direction of the quasar B0218+357 at $z=0.68466$
to yield a result that is consistent with no variation in $\mu $ when
systematic errors are more fully accounted for: $\Delta \mu /\mu =(+0.74\pm
0.47_{\mathrm{stat}}\pm 0.76_{\mathrm{system}})\times 10^{-6}$,
corresponding to a time variation of $\dot{\mu}/\mu =(-1.2\pm 0.8_{\mathrm{%
stat}}\pm 1.2_{\mathrm{system}})\times 10^{-16}$yr$^{-1\text{ }}$ in the
best fit $\Lambda CDM$ cosmology.

Any variation of $\alpha $ and $\mu $ today can also be constrained by
direct laboratory searches. These are performed by comparing clocks based on
different atomic frequency standards over a period of months or years. Until
very recently, the most stringent constraints on the temporal variation in $%
\alpha $ arose by combining measurements of the frequencies of Sr \cite%
{blatt}, Hg+ \cite{fortier}, Yb+ \cite{peiknew}, and H \cite{fischer} relative
to Caesium: $\dot{\alpha}/\alpha =(-3.3\pm 3.0)\times 10^{-16}\,\mathrm{yr}%
^{-1}$. Cingöz \emph{et al.} also recently reported a less stringent limit
of $\dot{\alpha}/{\alpha }=-(2.7\pm 2.6)\times 10^{-15}\,\mathrm{yr}^{-1}$ 
\cite{cingoz}; however, if the systematics can be fully understood, an
ultimate sensitivity of $10^{-18}\,\mathrm{yr}^{-1}$ is possible with their
method  \cite{nguyen}. If a linear variation in $\alpha $ is assumed then
the Murphy \emph{et. al.} quasar measurements equate to $\dot{\alpha}/{%
\alpha }=(6.4\pm 1.4)\times 10^{-16}\,\mathrm{yr}^{-1}$ \cite{webb}. If the
variation is due to a light scalar field described by a theory like that of
Bekenstein \cite{bek} and Sandvik, Barrow and Magueijo (BSBM) \cite{bsbm1},
then the rate of change in the constants is exponentially damped during the
recent dark-energy-dominated era of accelerated expansion, and one typically
predicts $\dot{\alpha}/\alpha =1.1\pm 0.3\times 10^{-16}\,\mathrm{yr}^{-1}$
from the Murphy \emph{et al} data, which is not ruled out by the atomic
clock constraints mentioned above. For comparison, the Oklo natural reactor
constraints, which reflect the need for the ${\rm Sm}^{149}+n\rightarrow
{\rm Sm}^{150}+\gamma $ neutron capture resonance at $97.3\,{\rm meV}$ to have been
present $1.8-2\,{\rm Gyr}$ ($z=0.15$) ago, as first pointed out by Shlyakhter \cite%
{sh}, are currently \cite{fuj} $\Delta \alpha /\alpha =(-0.8\pm 1.0)\times
10^{-8}$ or $(8.8\pm 0.7)\times 10^{-8}$ (because of the double-valued
character of the neutron capture cross-section with reactor temperature) and 
\cite{lam} $\Delta \alpha /\alpha >4.5\times 10^{-8}$ $(6\sigma )$ when the
non-thermal neutron spectrum is taken into account. However, there remain
significant environmental uncertainties regarding the reactor's early
history and the deductions of bounds on constants. The quoted Oklo constraints on $\alpha$ apply only when all other constants are held to be fixed.  If the quark masses to vary relative to the QCD scale, the ability of Oklo to constrain variations in $\alpha$ is greatly reduced \cite{Flambaun07}.

Recently, Rosenband \emph{et al.} \cite{Rosenband} measured the ratio of
aluminium and mercury single-ion optical clock frequencies, $f_{\mathrm{Al+}%
}/f_{\mathrm{Hg+}}$, repeated over a period of about a year. From these
measurements, the linear rate of change in this ratio was found to be $%
(-5.3\pm 7.9)\times 10^{-17}\,\mathrm{yr}^{-1}$. These measurements provides
the strongest limit yet on any temporal drift in the value of $\alpha $: $%
\dot{\alpha}/\alpha =(-1.6\pm 2.3)\times 10^{-17}\,\mathrm{yr}^{-1}$. This
limit is strong enough to strongly rule out theoretical explanations of the
change in $\alpha $ reported by Webb \emph{et al.} \cite{webb} in terms of
the slow variation of an effectively massless scalar field, even allowing
for the damping by cosmological acceleration, unless there is a significant
new physical effect that slows the locally observed effects of changing $%
\alpha $ on cosmological scales (for a detailed analysis of global-local
coupling of variations in constants see Refs. \cite{shawb}).

It has been noted that if the `constants' such as $\alpha $ or $\mu $ can
vary, then in addition to a slow temporal drift one would also expect to see
an annual modulation in their values. In many theories, the Sun perturbs the values of the constants by a factor roughly proportional to the
Sun's Newtonian gravitational potential \cite{bmag} (the contribution from
the Earth's gravitational potential is about 14 times smaller than that of
the Sun's at the Earth's surface). Hence the `constants' depend on the distance from
the Sun. Since the Earth's orbit around the Sun has a small ellipticity, the
distance, $r$, between the Earth and Sun fluctuates annually, reaching a
maximum at aphelion around the beginning of July and a minimum at perihelion
in early January. It was shown in Ref. \cite{seasonal} that in many varying
constant models, the values of the constants measured here on Earth, would
oscillate in a similar seasonal manner. Moreover, in many cases, this seasonal fluctuation is predicted to dominate  over any linear temporal drift \cite{seasonal}.

Specifically, let us suppose that the Sun creates a distance-dependent
perturbation to the measured value of a coupling constant, $\mathcal{C}$, of
amplitude $\delta \ln \mathcal{C}=C(r)$. If this coupling constant is
measured on the surface of another body (e.g. the Earth) which orbits the first
body along an elliptical path with semi-major axis $a$, period $T_{\mathrm{p}%
}$, and eccentricity $e\ll 1$, then to leading order in $e$, the annual
fluctuation in $\mathcal{C}$, $\delta \mathcal{C}_{\mathrm{annual}}$ will be
given by 
\begin{equation}
\frac{\delta \mathcal{C}_{\mathrm{annual}}}{\mathcal{C}}=-c_{\mathcal{C}%
}\cos \left( \frac{2\pi t}{T_{\mathrm{p}}}\right) +O(e^{2}),  \label{VarForm}
\end{equation}%
where $c_{\mathcal{C}}\equiv e\,a C^{\prime }(a)$, $C^{\prime }(a) = d C(r)/dr \vert_{r=a}$ and $t=nT_{\mathrm{p}}$,
for any integer $n$, corresponds to the moment of closest approach
(perihelion). In the case of the Earth moving around the Sun, over a period
of 6 months from perihelion to aphelion one would therefore measure a change
in the constant $\mathcal{C}$ equal to $2c_{\mathcal{C}}$. 
\begin{figure}[tbh]
\begin{center}
\includegraphics[scale=0.50]{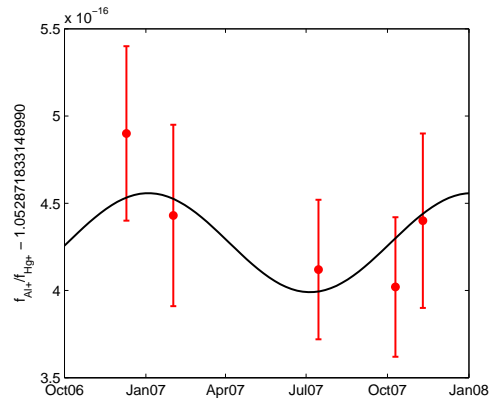}
\end{center}
\caption{(colour online). Frequency ratio $f_{\mathrm{Al+}}/f_{\mathrm{Hg+}}$
as measured by Rosenband \emph{et al.} \protect\cite{Rosenband}. The solid
black line shows the maximum likelihood fit for a seasonal variation.}
\label{fig1}
\end{figure}
Rosenband \emph{et al.} \cite{Rosenband} fitted a linear drift in $\alpha $
to their data finding $\dot{\alpha}/\alpha =(-1.6\pm 2.3)\times 10^{-17}\,%
\mathrm{yr}^{-1}$. We fit the expected form of any annual fluctuation, Eq. %
\ref{VarForm}, to the measured values of $f_{\mathrm{Al+}}/f_{\mathrm{Hg+}}$%
.  The data, taken from Ref. \cite{Rosenband}, and our best fit line, are
shown in Fig. \ref{fig1}. It should be noted that the magnitude of the systematic errors for the middle three data points were not verified to the same precision as they were from the first and last data points \cite{Rosenband}. We do not expect to this have a great effect on the magnitude of the resulting constraint on $k_{\alpha}$. Using $\delta \ln (f_{\mathrm{Al+}}/f_{\mathrm{Hg+}%
})=(3.19+0.008)\delta \alpha /\alpha $, \cite{Rosenband}, a maximum
likelihood fit to the data gives 
\begin{equation}
c_{\alpha }=ea\delta \alpha ^{\prime }(a)=\left( -0.89\pm 0.84\right) \times
10^{-17}.
\end{equation}%
where $a=149,597,887.5\,\mathrm{km}$ is the semi-major axis of the Earth's
orbit, and $\delta \alpha (r)$ is the perturbation in $\alpha $ due to the
Sun's gravitational field. Assume that over solar system scales, the values
of the scalar fields on which values of the 'constants' depend, vary with
the local gravitational potential \cite{bmag}. Hence, we have $\delta \alpha
(r)/\alpha =k_{\alpha }\Delta U_{\odot }(r),$ where $k_{\alpha \text{ }}$is
a theory-dependent multiplier, $\Delta U_{\odot }$ is the change in the
gravitational potential of the Sun: $U_{\odot }(r)=-GM_{\odot }/r$, and so $%
ea\Delta U_{\odot }^{\prime }(a)=eGM_{\odot }/a=1.65\times 10^{-10}$.
Hence, we find:
\begin{equation}
k_{\alpha }=(-5.4\pm 5.1)\times 10^{-8}.  \label{kalpha}
\end{equation}%
The frequency shifts measured by Rosenband \emph{et al.} \cite{Rosenband}
were not sensitive to changes in the electron-proton mass ratio: $\mu =m_{%
\mathrm{e}}/m_{\mathrm{p}}$. Measurements of optical transition frequencies
relative to Cs, Refs. \cite{blatt, fortier, peiknew, fischer}, are sensitive
to both $\mu $ and $\alpha $. H-maser atomic clocks \cite{Ashby} are also
sensitive to variations in the light quark to proton mass ratio: $q=m_{%
\mathrm{q}}/m_{\mathrm{p}}$. We can use all these observations if we define
two more gravitational coupling multipliers, $k_{\mu }$ and $k_{\mathrm{q}}$%
, by $\delta \mu /\mu =k_{\mu }\Delta U_{\odot }$, and $\delta q/q=k_{%
\mathrm{q}}\Delta U_{\odot }$. Refs. \cite{blatt, fortier,Ashby} give $%
k_{\alpha }+0.36k_{\mu }=(-2.1\pm 3.2)\times 10^{-6}$, $k_{\alpha
}+0.17k_{\mu }=(3.5\pm 6.0)\times 10^{-7}$, and $k_{\alpha }+0.13k_{\mathrm{q%
}}=(1\pm 17)\times 10^{-7}$ respectively. We also performed a bootstrap
seasonal fluctuation fit (with $10^{5}$ resamplings) to the $\mathrm{Yb}^{+}$
frequency measurements of Peik \emph{et al.} \cite{peiknew, peikprivate}
giving $k_{\alpha }+0.51k_{\mu }=(7.1\pm 3.4)\times 10^{-6}$. Combining
these bounds with Eq. (\ref{kalpha}) gives: 
\begin{eqnarray}
k_{\mu } &=&(3.9\pm 3.1)\times 10^{-6},  \label{kmu} \\
k_{\mathrm{q}} &=&(0.1\pm 1.4)\times 10^{-5}.  \label{kq}
\end{eqnarray}

Recently, Blatt \emph{et al.} \cite{blatt} combined data from measurements
of H-maser \cite{Ashby} and optical atomic clocks \cite{blatt,fortier}, to
bound the multipliers, $k_{\alpha }$, $k_{\mu }$ and $k_{q}$, finding: 
\begin{eqnarray}
k_{\alpha } &=&(2.5\pm 3.1)\times 10^{-6},  \notag \\
k_{\mu } &=&(-1.3\pm 1.7)\times 10^{-5},  \notag \\
k_{\mathrm{q}} &=&(-1.9\pm 2.7)\times 10^{-5}.  \notag
\end{eqnarray}%
The constraint on $k_{\alpha }$ derived in this paper from the data of
Rosenband \emph{et al.} \cite{Rosenband} therefore represents an improvement
by about two orders of magnitude over the previous best bound. This improved
bound on $k_{\alpha }$ combined with data found by Peik \emph{et al.} \cite%
{peiknew,peikprivate} has also produced an order of magnitude improvement in
the determination of $k_{\mu }$ and a slight improvement in the constraint
on $k_{q}$.

Seasonal fluctuations are predicted by a varying constant theory because the
scalar field which drives the variation in the constant couples to normal
matter. The presence of the Sun therefore induces gradients in scalar field
and the 'constants', and it is essentially these gradients that are
detectable as seasonal variables. However, as is well known, gradients in a
scalar field which couples to normal matter result in new or 'fifth' forces
with pseudo-gravitational effects. In the case of varying $\alpha $ and $\mu 
$ theories, these forces are almost always composition dependent, which
would violate the universality of free-fall and hence the weak equivalence
principle (WEP). The magnitude of any composition-dependent fifth force
toward the Sun is currently constrained to be no stronger than $%
10^{-12}-10^{-13}$ times than the gravitational force \cite{WEP}. In the
context of a given theory the constraints from WEP tests indirectly bound $%
k_{\alpha }$. Indeed, they often provide the tightest constraints on $%
k_{\alpha }$ \cite{seasonal,dent, bmag2}.

A recent and thorough analysis of the WEP violation constraints on $%
k_{\alpha }$ \cite{dent,dentprivate} found: 
\begin{equation*}
k_{\alpha }=(0.3\pm 1.7)\times 10^{-9},
\end{equation*}%
with a similar constraint on $k_{q}$. It must be noted, however, that this
result is still subject to theoretical uncertainty, especially regarding the
dependence of nuclear properties on quark masses. For instance, it was also
noted in Ref. \cite{dent} that if certain (fairly reasonable) assumptions
about nuclear structure are dropped, the 1$\sigma $ error bars on $k_{\alpha
}$ increase by about an order of magnitude to: $\pm 1.4\times 10^{-8}$.
Despite these uncertainties, for many
theories of varying $\alpha $, WEP violation constraints from laboratory experiments or lunar laser ranging  \cite{nord} still provide the
strongest, albeit indirect, bound on $k_{\alpha }$.

We have shown above that direct constraints on any change in $\alpha $ that
is proportional to the local gravitational potential (as in general theories
of its spacetime variation \cite{bmag}) are now within less than $1.5$ orders of magnitude
of those extrapolated, under certain fairly reasonable assumptions, from the
non-detection of any WEP violation. One might wonder how much further the
sensitivity of atomic clock experiments to variations in $\alpha $ would
have to improve before direct constraints on $k_{\alpha }$ would surpass
those coming from the current WEP violation bounds. Suppose that over a few
days, the ratio of two transition frequencies, $f_{\mathrm{A}}/f_{\mathrm{B}}
$, can be measured with a sensitivity $\sigma _{\mathrm{f}}$, and that $%
\delta \ln (f_{\mathrm{A}}/f_{\mathrm{B}})=S_{\alpha }\delta \alpha /\alpha $
(typically $S_{\alpha }\sim \mathcal{O}(1),$ although some transitions
exhibit a greatly increased sensitivity \cite{flamth}). The sensitivity to
changes in $\alpha $ is then given by $\sigma _{\alpha }=\sigma _{\mathrm{f}%
}/S_{\alpha }$. By simulating data sets, we found that the sensitivity to $%
k_{\alpha }$ is significantly improved if one makes $N_{\mathrm{m}}\gtrsim 12
$ measurements per year (at roughly regular intervals). With $N_{\mathrm{m}%
}\gtrsim 12$, by performing a bootstrap linear regression with $10^{5}$
resamplings of the simulated data points, we find that the sensitivity, $%
\sigma _{k}$ to $k_{\alpha }$ is roughly: 
\begin{equation*}
\sigma _{\mathrm{k}}\approx 0.69\times 10^{10}\frac{\sigma _{\alpha }}{\sqrt{%
N_{\mathrm{y}}(N_{\mathrm{m}}-1)}}.
\end{equation*}%
where $N_{\mathrm{y}}$ is the number of years for which data is taken. The
total number of measurements is therefore $N_{\mathrm{y}}N_{\mathrm{m}}$.
Indirect constraints currently have a sensitivity of $\sigma _{\mathrm{k}%
}=1.7\times 10^{-9}$ \cite{dent}. This would be surpassed by direct
measurements if $\sigma _{\alpha }<2.5\sqrt{N_{\mathrm{y}}(N_{\mathrm{m}}-1)}%
\times 10^{-19}$. For example, if we take measurements every 20 days (or so)
over a single year ($N_{\mathrm{m}}=18$, $N_{\mathrm{y}}=1$) then we would
require $\sigma _{\alpha }\lesssim 10^{-18}$. The experiment conducted by
Rosenband \emph{et al.} \cite{Rosenband} currently has a $\sigma _{\mathrm{f}%
}\approx 5\times 10^{-17}$ and $S_{\alpha }=3.2$, so $\sigma _{\alpha
}\approx 1.6\times 10^{-17}$; one would therefore require an unrealistic
number of measurements $N_{\mathrm{y}}(N_{\mathrm{m}}-1)\geq 4100$ to
achieve the desired sensitivity to $k_{\alpha }$. Cingöz \emph{et al.}
obtained constraints on $k_{\alpha }$ by monitoring the transition
frequencies between nearly degenerate, opposite-parity levels in two
isotopes of atomic Dysprosium \cite{cingoz}. Currently sensitively to frequency changes is $7.5\mathrm{Hz}$, giving $\sigma _{\alpha }=5\times 10^{-15}$. However, an ultimate
 sensitivity of about a mHz is predicted to be achievable \cite{nguyen}. This would give $\sigma
_{\alpha }\approx 7\times 10^{-19}$ and one would
only require $N_{\mathrm{y}}(N_{\mathrm{m}}-1)\gtrsim 8$ to surpass the
current WEP violation constraints on $k_{\alpha }$. In addition, Flambaum
noted in Ref. \cite{flamth} that the transition between the ground and the
first excited states in the ${}^{229}\mathrm{Th}$ nucleus is particularly
sensitive to changes in $\alpha $ and $\mu $ with $S_{\alpha }\sim 10^{5}$.
By measuring this transition frequency relative to, say, a Cs frequency
standard, and without any improvement in measurement accuracy over what has
currently been achieved, it may be possible bring $\sigma _{\alpha }$ down
to about $10^{-20}$, giving $\sigma _{\mathrm{k}}\lesssim 2\times 10^{-11}$
with $N_{\mathrm{m}}\gtrsim 12$ -- two orders of magnitude better than the
current WEP violation constraints. The motivation for a future space-based
test of the WEP with possible sensitivity of order $10^{-15} - 10^{-18}$ therefore
remains very great \cite{microscope,step}.

In summary: we have shown how new laboratory constraints on possible time
variation in the fine structure `constant' and the electron-proton mass
ratio can yield more sensitive limits by incorporating the effects of the
seasonal variation of the Sun's gravitational field at the Earth's surface.
This seasonal variation is expected in all theories which require that the
covariant d'Alembertian of any scalar field driving variation of a
`constant' is proportional to the dominant local source of gravitational
potential \cite{bmag2}. The recent experimental results from Rosenband et al 
\cite{Rosenband} and Peik et al, \cite{peiknew} \ have reached the
sensitivity of the quasar observations of varying $\alpha $ and $\mu $ made
at high redshift and we have shown may soon provide stronger bounds on
varying constants than conventional ground-based WEP experiments.

\vspace{0.5cm}

\noindent \textbf{Acknowledgements:} DJS is supported by STFC. JDB would
like to thank Paolo Molaro for discussions and for drawing attention to Ref. 
\cite{Rosenband}. We are grateful to Thomas Dent and Ekkehard Peik for
helpful discussions and for providing additional information regarding Refs. 
\cite{dent} and \cite{peiknew} respectively.

\end{document}